\documentclass[twocolumn]{article}
\usepackage{amssymb}
\usepackage{xy} \xyoption{all}

\newtheorem{remark}{Remark}[section]
\newtheorem{proposition}{Proposition}[section]

\newcommand{\R}{\mathbb{R}}
\newcommand{\Z}{\mathbb{Z}}
\newcommand{\id}[1]{\mbox{\rm Id}_{#1}}
\newcommand{\pint}[2]{\langle #1|#2 \rangle}
\newcommand{\pr}[1]{\mbox{\rm Prb}\left[#1\right]}
\newcommand{\casos}[4]{\left\{\begin{array}{ll}
#1 &\mbox{ #2, } \\
#3 &\mbox{ #4. } 
\end{array}\right.}
\newcommand{\prob}[3]{
\noindent{\bf Problem} {\tt #1}
\vspace{2ex}
 
\noindent {\bf Instance:} \begin{minipage}{20em}#2\end{minipage}
\vspace{1ex}

\noindent {\bf Solution:} \ \begin{minipage}{20em}#3\end{minipage}
\vspace{2ex}}

\title{
A Geometric Presentation of Probabilistic Satisfiability
}
\author{Guillermo Morales-Luna \\
Computer Science Department \\
CINVESTAV-IPN \\
Mexico City, Mexico \\
{\tt gmorales@cs.cinvestav.mx}
}
\begin{document}
\maketitle

\begin{abstract}
By considering probability distributions over the set of assignments the expected truth values assignment to propositional variables are extended through linear operators, and the expected truth values of the clauses at any given conjunctive form are also extended through linear maps. The probabilistic satisfiability problems are discussed in terms of the introduced linear extensions. The case of multiple truth values is also discussed.
\end{abstract}

\section{Introduction}

{\tt SAT}, the {\em satisfiability} problem for Boolean forms, is the typical NP-complete problem and many variants of it have been widely studied~\cite{BH09}. {\em Probabilistic Satisfiability}, {\tt PSAT}, considers probability distributions on the collection of assignments and for a given conjunctive form and an expected sample truth values for the clauses appearing in the form it asks whether there is a probability distributions on the assignments realizing that expected sample of values. {\tt PSAT} was formally introduced in the 80's~\cite{Nil86} and it has been proved NP-complete~\cite{GK88}. Due to the linear nature of the expectation operator, a presentation based on linear maps is well suited for {\tt PSAT}. Here we follow already canonical expositions of the problem~\cite{Han99,Pr05} and we pose the problem through linear real spaces.

The outline of the current exposition is the following: In section~\ref{sc.vras} we introduce the basic notions related to assignments and probability distributions over them, in section~\ref{sc.ext} we see that assignments to variables may be extended by linear maps which act as the linear Hamming codes over real spaces, we pose the {\em Coherence Problem} and we establish a geometrical condition for two distributions determining the same expected variable assignment. In section~\ref{sc.cfs} we deal with the expected truth values of conjunctive forms and we pose {\tt PSAT}. The reduction of {\tt SAT} to {\tt PSAT} is discussed in geometrical terms. In section~\ref{sc.vars} some variants of {\tt PSAT} are posed as linear integer optimization problems. Finally, in section~\ref{sc.mvl} our approach is put in more general terms in order to cover the case of a multivalued logic, in line with the exposition in~\cite{Qi03}.

In the paper we denote by $[\![i,j]\!]$ the set of integers $\{i,i+1,\ldots,j-1,j\}$, with $i,j\in\Z$, $i\leq j$.

\section{Boolean variables and assignments}\label{sc.vras}

Let $X=\left(X_j\right)_{j=0}^{n-1}$ be a set of propositional variables and let $B^n=\{0,1\}^n$ be the assignment space, namely, the $n$ dimensional hypercube.

The $2^n$ assigments are enumerated in a canonical way: For each $j\in[\![0,2^n-1]\!]$ the assignment $\sigma_j$ is obtained by expressing $j$ in base-2 as an $n$-length bit string.

Let $U^n= [0,1]^n$ be the unit cube in the real space $\R^n$, its vertexes are the points at the hypercube $B^n$. Let $\Pi_n$ be the collection of probability distributions over the hypercube,
$P:B^n\to[0,1]$, $\sigma\mapsto P(\sigma)$, with $\sum_{\sigma\in B^n} P(\sigma) = 1$. Clearly, $\Pi_n$ is included in the affine hyperplane that contains the points at the canonical basis:
$$D_{2^n} = e_0+\left({\bf 1}_{2^n}\right)^{\perp}.$$
For each distribution $P\in\Pi_n$, let $u_P=\left(P(\sigma)\right)_{\sigma\in B^n}\in U^{2^n}$ be the vector whose entries are the probability values. The expected value of the assignments under $P$, $x_P=\sum_{\sigma\in B^n} P(\sigma) \sigma$, is a point in the unit cube $U^n$. Let us write it as
$$x_P=\left(x_{Pj}\right)_{j=0}^{n-1}.$$
For each index $i\in[\![0,n-1]\!]$, let
\begin{equation}
z_{Pi} = \sum_{\sigma_i=1}P(\sigma) - \sum_{\sigma_i=0}P(\sigma) = 2\sum_{\sigma_i=1}P(\sigma)-1.\label{eq.010}
\end{equation}
Then, $z_{Pi}>0$ whenever $1$ is the most probable value for the $i$-propositional variable $X_i$.
Let $z_P=\left(z_{Pi}\right)_{i=0}^{n-1}$.

For each $i\in[\![0,n-1]\!]$, we have indeed $\pr{X_i=\mbox{\tt True}}=x_{Pi}$. Thus the vector $x_P\in U^n$ is a probabilistic assignment to the propositional variables.
Reciprocally, a vector $x\in U^n$ is called {\em coherent} if there exists a $P\in\Pi_n$ such that $x=x_P$.

\section{Linear extensions of assignments}\label{sc.ext}

For any distribution $P\in\Pi_n$, let $u_P=\left(P(\sigma_j)\right)_{j=0}^{2^n-1}$ be the $2^n$-dimensional vector whose entries are the probability values. Let us observe that the map $u_P\mapsto x_P$ introduced above is the restriction of the linear map $W_n:\R^{2^n}\to\R^n$ represented by the matrix of order $n\times 2^n$ whose columns are the elements at the hypercube $B^n$. By ordering these points according to their Hamming weights, the matrix $W_n$ can be written as
\begin{equation}
W_n = [{\bf 0}_n\ \ \id{n}\ \ H_n],\label{eq.0100}
\end{equation}
where ${\bf 0}_n$ is the zero vector in $\R^n$, $\id{n}$ is the identity matrix of order $n\times n$ and $H_n$ is a real matrix (indeed with entries 0, 1) of order $n\times(2^n-n-1)$. Thus $W_n$ is a full-rank matrix, its image has dimension $n$ and its orthogonal complement in $\R^{2^n}$ has dimension $2^n-n$. Indeed, a generator matrix of the orthogonal complement is
\begin{equation}
K_n = \left[\begin{array}{ll}
1 & {\bf 0}_{2^n-n-1}^T \\
{\bf 0}_n & -H_n \\
{\bf 0}_{2^n-n-1} & \id{2^n-n-1}
\end{array}\right] \in B^{2^n\times(2^n-n)}.\label{eq.011}
\end{equation}
Consequently, 
$$\forall u_0,u_1\in\R^{2^n}\,\left[W_nu_0=W_nu_1\ \Leftrightarrow\ u_1-u_0\in K_n\right].$$
Let us state the following:

\prob{Coherence}
{A vector $x\in U^n$.}
{$\casos
{1}{if $\exists u\in\R^{2^n}$: 
$\begin{array}[t]{rcl}
x&=&W_nu \\
\pint{{\bf 1}_{2^n}}{u}&=&1\\
u&\geq&{\bf 0}_{2^n}
\end{array}$
}
{0}{otherwise}$
}
(${\bf 0}_{2^n}$ and ${\bf 1}_{2^n}$ are the constant vectors 0 and 1 of dimension $2^n$.)

Also, it is worth to mention that the correspondence $u_P\mapsto z_P$ defined by the relation~(\ref{eq.010}) is the restriction to $\Pi_n$ of the linear map $Z_n = 2W_n-{\bf 1}_{n\times 2^n}:\R^{2^n}\to\R^n$.
If $x\in U^n$ is coherent and $x=W_nu$ then
$$Z_nu = 2x - \left(\sum_{j=0}^{2^n-1}u_j\right)\,{\bf 1}_n = 2x - {\bf 1}_n.$$
\begin{remark}
$x\in U^n$ is coherent if and only if $u\in\R^{2^n}$ is such that
$$2x - {\bf 1}_n=Z_nu\ ,\ \pint{{\bf 1}_{2^n}}{u}=1\ ,\ u\geq{\bf 0}_{2^n}.$$
   \end{remark}

Now let us assume that for a vector $x\in U^n$ there are two distributions $u_0,u_1\in\Pi_n$ such that $W_nu_0 = x = W_nu_1$. Then $v=u_1-u_0\in K_n$, hence there exists a vector $w\in\R^{2^n-n}$ such that $v=K_nw$. Since the components of $u_0$ y $u_1$ have as addition 1, necessarily $\pint{c_n}{w}=0$ where 
$$c_n = [1]\star\left(\bigstar_{i=1}^{n-1}[(-i)^{{n\choose i+1}}]\right)\in\R^{2^n-n}$$
(here the operator $\star$ is list concatenation). 

For each distribution $u_0\in\Pi_n$ let
$$T_n(u_0) = \{w\in c_n^{\perp}\subset\R^{2^n-n}|\ K_nw+u_0 \geq {\bf 0}_n\}.$$
This is a polyhedron contained in a $(2^n-n-1)$-dimensional linear subspace of $\R^{2^n-n}$, thus $K_n(T_n(u_0))$ is a $(2^n-n-1)$-dimensional polyhedron in $\R^{2^n}$.

\begin{remark}
If $u_0$ realizes the coherence of a probabilistic assignment $x\in U^n$ then any distribution $v+u_0$, with $v\in K_n(T_n(u_0))$, also realizes the coherence.
   \end{remark}

\section{Conjunctive forms}\label{sc.cfs}

Let $F=\left(C_i\right)_{i=0}^{m-1}$ be a conjunctive form consisting of $m$ clauses over the set $X$ of $n$ propositional variables.

For each assignment $\sigma:X\to B$ let $v_{\sigma}=\left(\sigma(C_i)\right)_{i=0}^{m-1}\in\{0,1\}^m$ be the vector of truth values corresponding to the clauses in $F$ under $\sigma$. Let $V_{mn}=\left[v_{\sigma}\right]_{\sigma\in B^n}\in B^{m\times 2^n}$ be the matrix whose columns are the vectors $v_{\sigma}$, with $\sigma\in B^n$. Then $\forall (i,j)\in[\![0,m-1]\!]\times[\![0,2^n-1]\!]$:
\begin{equation}
\ \left[v_{ij}=1 \Longleftrightarrow \sigma_j(C_i)=1\right]. \label{eq.01}
\end{equation}
Thus, $\forall u\in\R^{2^n}\,\forall i\in[\![0,m-1]\!]$: 
\begin{equation}
(V_{mn}u)_i = \sum\{u_{ij}|\ \sigma_j(C_i)=1\}. \label{eq.02}
\end{equation}
Naturally, if there is an index $i$, such that $(V_{mn}u)_i \not= 0$ then there exists an assignment $\sigma_j\in B^n$ such that $u_{ij} \not= 0$ and $\sigma_j$ satisfies the clause $C_i$. 

Besides, if $u\in\Pi_n$ is a distribucition and $(V_{mn}u)_i = 1$ then the support of $u$, $\mbox{Spt}(u)=u^{-1}(1)$, is contained in the support of the $i$-th row of $V_{mn}$: $\forall i\in[\![0,m-1]\!]$
\begin{equation}
\begin{array}{rl}
&\left[(V_{mn}u)_i = 1\ \Longrightarrow\ \right. \\
&\left.\forall j\in[\![0,2^n-1]\!]\ \left[u_j\not=0\Longrightarrow\sigma_j(C_i)=1\right]\right].
\end{array} \label{eq.021}
\end{equation}
Let $P\in\Pi_n$ be a probability distribution over $B^n$. For each clause $C_i$, let 
$$y_{Pi}=\sum\{P(\sigma) |\ \sigma(C_i)=1\}.$$
Thus, with respect to $P$, $\pr{C_i=\mbox{\tt True}}=y_{Pi}$. Hence, if $P$ ``is more concentrated'' on the assignments satisfying $C_i$ then ``it would be more probable for $C_i$ to be true under $P$''.

The expected truth values vector $y_P=\left(y_{Pi}\right)_{i=0}^{m-1}\in U^m$ is indeed determined as $y_P=V_{mn}u_P$. Thus expectation is the restriction of the linear map $V_{mn}:\R^{2^n}\to\R^m$, $u\mapsto y=V_{mn}u$. 

The above introduced linear maps are shown in the following diagram:
\begin{equation}
\xymatrix{
 & \R^n  \\
\R^{2^n} \ar[r]^{W_n} \ar[dr]_{V_{mn}} \ar[ur]^{Z_n} & \R^n \ar@{..>}[d]^{E_0} \\
 & \R^m  %
}\label{eq.d01}
\end{equation}
And their restrictions to the assignment spaces produce the diagram:
\begin{equation}
\xymatrix{
\Pi_n \ar[r]^{W_n} \ar[dr]_{V_{mn}} & U^n \ar@{..>}[d]^{(E_0)_{U_n}} \ar@/^/[r]^{\rho} & B^n \ar@/^/[l]^{\iota} \ar[d]^{E} \\
 & U^m \ar@/^/[r]^{\rho} & B^m \ar@/^/[l]^{\iota} %
}\label{eq.d02}
\end{equation}
where the maps $\iota$ are simple embeddings, $E:B^n\to B^m$ is the truth evaluation map of clauses over Boolean assignments, and $\rho$ is a ``determination" operator, e.g.
$$x=\left(x_i\right)_i\ \Longrightarrow\ \rho(x)=y=\left(y_i\right)_i$$
with
$$y_i = \left\{\begin{array}{cl}
1 & \mbox{ if }x_i\geq\frac{1}{2} \vspace{1ex}\\
0 & \mbox{ if }x_i<\frac{1}{2} 
\end{array}\right.$$
Nevertheless:

\begin{remark}
In general, there is no a map $E_0:\R^n\to\R^m$ such that $(E_0)_{U_n}$ makes the diagram~(\ref{eq.d02}) commutative.
\end{remark}

Namely, if there is such $E_0$ then $E_0\circ W_n=V_{mn}$, and $\forall u\in\R^n$, $\forall v\in K_n$:
\begin{eqnarray*}
V_{mn}(u)+V_{mn}(v) &=& V_{mn}(u+v) \\
 &=& E_0\circ W_n(u+v) \\
 &=& E_0\circ W_n(u) \\
 &=&V_{mn}(u) 
\end{eqnarray*}
or equivalently
\begin{equation}
\mbox{ker}(W_n) = K_n\subseteq \mbox{ker}(V_{mn}).\label{eq.041}
\end{equation}
From relation~(\ref{eq.011}) we observe that $e_0=\{1\}\star{\bf 0}_{2^n-1}\in K_n$, 
but if the first clause $C_0$ of $F$ is satisfied by the assignment ${\bf 0}_{2^n}$ then~(\ref{eq.041}) cannot hold. \hfill$\Box$
\medskip

Let $\Gamma_m=V_{mn}(\Pi_n)\subset\R^m$ be the image of $\Pi_n$ under the map $P\mapsto V_{mn}u_P$. The decision problem for $\Gamma_m$ is posed equivalently as {\em Probabilistic Satisfiability}:

\prob{SATP}
{A vector $y\in U^m$.}
{$\casos
{1}{if $\exists u\in\R^{2^n}$: 
$\begin{array}[t]{rcl}
y&=&V_{mn}u \\
\pint{{\bf 1}_{2^n}}{u}&=&1\\
u&\geq&{\bf 0}_{2^n}
\end{array}$
}
{0}{otherwise}$}

In particular, for the instance $y={\bf 1}_m$, if the answer of {\tt SATP} is 1 and $u\in\R^{2^n}$ is the witnessing distribution, then from the relation~(\ref{eq.021}),
\begin{equation}
\mbox{Spt}(u)\subset\bigcap_{i=0}^{m-1}\mbox{Spt}(C_i). \label{eq.031}
\end{equation}
Since the intersection at the right side of~(\ref{eq.031}) is not empty, then there is a classical assignment satisfying the whole conjunctive form $F$. This is just a restatement of the well known

\begin{proposition}\label{pr.dif} {\tt SATP} is NP-hard.
\end{proposition}

\section{Some variants}\label{sc.vars}

Let $F=\left(C_i\right)_{i=0}^{m-1}$ be a conjunctive form, and let $C_m$ be another clause. Let $z_m=\left(\sigma(C_m)\right)_{\sigma\in B^n}\in B^{2^n}$ be the vector consisting of the truth values of the clause $C_m$ over all the assignments. The {\em Entailment Probabilistic Satisfiability} problem is the following:

\noindent{\tt EPSAT}
\begin{eqnarray*}
\mbox{Minimize } & & \pint{z_m}{u}  \\
\mbox{subject to } & & \pint{{\bf 1}_{2^n}}{u}=1\ ,\ u\geq{\bf 0}_{2^n} 
\end{eqnarray*}

If $u\in\R^{2^n}$ is a solution of {\tt EPSAT} and $a_m=\pint{z_m}{u}$ then a solution of {\tt PSAT}$(y)$ can be extended to a solution of {\tt PSAT}$(y*\{y_m\})$ whenever $0\leq y_m\leq a_m$. 

{\tt PSAT} and {\tt EPSAT} can be posed as sbproblems of the following:

\noindent{\tt OptPSAT}: Given $V_{mn}\in B^{m\times 2^n}$, $z\in B^{2^n}$ and $a,b\in\R^m$:
\begin{eqnarray*}
\mbox{Minimize } & & \pint{z}{u}  \\
\mbox{subject to } & & a\leq V_{mn}u\leq b\ ,\ \pint{{\bf 1}_{2^n}}{u}=1\ ,\ u\geq{\bf 0}_{2^n} 
\end{eqnarray*}

Evidently, {\tt OptPSAT} is an integer linear programming thus it can be reduced to {\tt SAT}. Hence, together with the proposition~\ref{pr.dif}:

\begin{proposition} PSATP is NP-complete.
\end{proposition}

\section{Multivalued logic}\label{sc.mvl}

Let $k\geq 2$ be an integer and let $B_k=\left\{\frac{\kappa}{k-1}\right\}_{\kappa=0}^{k-1}$ be the set of $k$ truth values, the minimum, $0$, corresponds t the value {\em False} while the maximum, $1$, is the truth value {\em True}. Let $X=\left\{X_j\right\}_{j=0}^{n-1}$ be a set of $n$ propositional variables. The assignment space $B_k^n$ possesses $k^n$ elements. Let $\Pi_{nk}$ be the collection of probability distributions over $B_k^n$. As in the relation~(\ref{eq.010}), the expected assignment, with respect to a distribution $u\in B_k^n$, is the restriction to $\Pi_{nk}$ of the linear map $\R^{k^n}\to\R^n$ determined by the matrix
\begin{equation}
W_{nk} = \frac{1}{k-1} [{\bf 0}_n\ \ \id{n}\ \ H_{nk}]\in B_k^{n\times k^n}, \label{eq.11}
\end{equation}
whose columns are the $k^n$ assignments sorted as follows: for each integer $\nu\in[\![0,n-1]\!]$ and for each string $\sigma$ consisting of $\nu$ digits $\frac{\kappa}{k-1}$ with $\kappa\not=0$, there appear the ${n\choose\nu}$ possibilities to put $\sigma$ in an assignment with Hamming weight $\nu$.

A generator matrix of the orthogonal complement of the image of $W_{nk}$ is given, as in relation~(\ref{eq.011}), as
\begin{equation}
K_{nk} = \frac{1}{k-1} \left[\begin{array}{ll}
1 & {\bf 0}_{k^n-n-1}^T \\
{\bf 0}_n & -H_{nk} \\
{\bf 0}_{k^n-n-1} & \id{k^n-n-1}
\end{array}\right].\label{eq.12}
\end{equation}
Clearly, $K_{nk}  \in B^{k^n\times(k^n-n)}$.

The problem {\tt Coherence} is posed also within this context.

Now, for the Boolean connectives let us consider the following propagation maps:
\begin{eqnarray*}
\neg:\kappa&\mapsto& 1 - \kappa \\
\lor:(\kappa_0,\kappa_1)&\mapsto&\max(\kappa_0,\kappa_1) \\
\land:(\kappa_0,\kappa_1)&\mapsto&\min(\kappa_0,\kappa_1).
\end{eqnarray*}
For any conjunctive form $F=\left(C_i\right)_{i=0}^{m-1}$, let $V_{nmk} = \left[v_{ij}\right]_{0\leq i\leq m-1}^{0\leq j\leq k^n-1}$ be the matrix such that at each entry $ij$ it has the value of the $i$-th clause corresponding to the $j$-th assignment: $$v_{ij} = C_i(\sigma_j).$$
The matrix $V_{nmk}$ determines a linear map $V_{nmk}:\R^{k^n}\to\R^m$. For any probability distribution $u\in\Pi_{nk}$, the image $y=V_{nmk}u$ gives the expected truth values for the clauses according to $u$. Consequently PSAT can also be posed as well within this context, let us call it {\tt PSAT}$_k$. 

For the instance ${\bf 1}_m\in U^m$, there exists a solution of {\tt PSAT}$_k$ para $F$ if and only if the conjunctive form $F$ is satisfiable in the classical sense.


\end{document}